\documentclass[twocolumn,prl,aps,showpacs]{revtex4}
\usepackage{amssymb}
\usepackage{epsfig}
\usepackage{amsbsy}
\usepackage{amsmath}
\usepackage{graphicx}

\begin{document}

\title{Supersymmetric Response of Bose-Fermi Mixture to Photoassociation }
\author{T. Shi}
\author{Yue Yu}
\author{C. P. Sun }
\affiliation{Institute of Theoretical Physics, Chinese Academy of Sciences, P.O. Box
2735, Beijing 100190, China}
\date{\today}

\begin{abstract}
 We study supersymmetric (SUSY) responses to a
photoassociation(PA) process in a mixture of Bose molecules $b$
and Fermi atoms $f$ which turn to mutual superpartners for a set
of proper parameters. We consider the molecule $b$ to be a bound
state of the atom $f$ and another Fermi atom $F$ with different
species. The $b$-$f$ mixture and a free $F$ atom gas are loaded in
an optical lattice. The SUSY nature of the mixture can be signaled
in the response to a photon-induced atom-molecule transition:
While two new types of fermionic excitations, an individual $b$
particle-$f$ hole pair continuum and the
Nambu-Goldstone-fermion-like ( or goldstino-like) collective mode,
are concomitant for a generic $b$-$f$ mixture, the former is
completely suppressed in the SUSY $b$-$f$ mixture and the
zero-momentum mode of the latter approaches to an exact
eigenstate. This SUSY response can be detected by means of the
spectroscopy method, e.g., the
PA spectrum which displays the molecular formation rate of $%
Ff\to b$.

\end{abstract}


\maketitle

\noindent \textit{Introduction --- } Recently, studies in the
supersymmetry (SUSY) for a mixture of cold Bose and Fermi atoms
have made spectacular progress \cite{th2,th1,yu}. In such a cold
atomic system, however, a Bose atom never transits to a Fermi
atom, its superpartner or vice verse. In addition to the
nonrelativity, this is another essential difference of this
low-energy SUSY from the SUSY in high-energy physics. For the
latter, such SUSY decay processes are always anticipated, e.g., a
quark (lepton) may emit or absorb a gaugino and decays to a squark
(slepton), the superpartner of the quark (lepton)~\cite{Weinberg}.

To expose the interesting SUSY nature of the mixture, the
effective "decay" process must be introduced. For a cold atomic
SUSY mixture with Bose-Einstein condensation, there is an
effective decay of SUSY generators
since they behave as the fermion annihilation and creation operators \cite%
{yu}. Therefore, the SUSY excitations can be simulated by a
boson-enhanced fermionic excitation. As a result, a
Nambu-Goldstone-fermion-like (or "goldstino-like")  mode in the
condensation phase of bosons could be observed by means of the
single-particle spectroscopy~\cite{SRS,PS}.

To achieve an exact SUSY mixture, the system parameters have to be
fine-tuned, which requires elaborate experimental setups and then
loses the generality.  In this article, we explore how to observe
the SUSY response by means of a spectroscopy measurement, even if
the mixture deviates slightly from the SUSY and the bosons do not
condense to form a whole ordered phase. This can resolve the
fine-tuning restraints in measuring the SUSY response. On the
other hand, the explicit breaking of the SUSY may create new
excitations, the bosonic particle-fermionic hole individual
continuous excitations, other than the collective goldtino-like
mode. Although our theory is nonrelativistic, the creation of
these new excitations due to SUSY explicit breaking should be
quite general . This may be a helpful point to the study of SUSY
in relativistic theory.

\begin{figure}[tbp]
\includegraphics[bb=41 374 562 767, width=8.5 cm, clip]{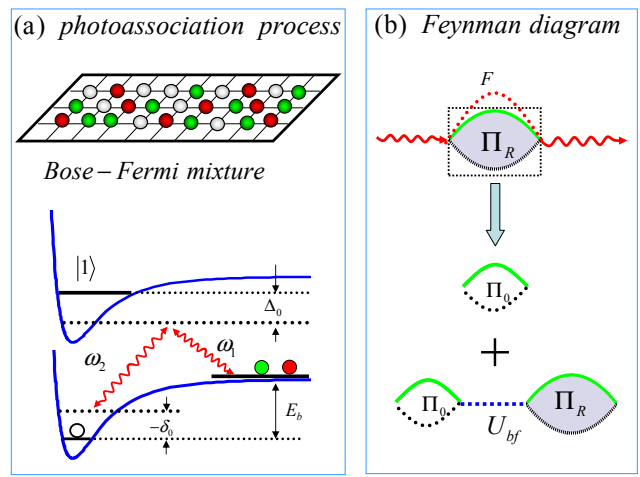}
\caption{(Color online) (a) Up-panel: The optical lattice with
cold particles. The gray (green), black (red), and white dots
denote Fermi atoms $f$, $F$, and the molecule $b$, respectively.
Low panel: The PA processes of two atoms to one molecule with the
binding energy $E_{b}$. (b) The Feynman diagram for the linear
response theory: The Green function of $Q$ is calculated by RPA.
Up-panel: The wavy (red) lines and dotted (red) line denote the
free Green functions of photon and $F$, respectively. Low panel:
The solid (green) curves and dotted (black) curves denote the free
Green functions of $f$ and $b$, respectively.} \label{fig1}

\end{figure}

We consider a mixture of Bose molecules $b$ and Fermi atoms $f$
with on-site interaction in a $d$-dimensional optical lattice
$(d=2,3)$ (see Fig.1(a)). With properly tuned interactions and
hopping amplitudes, this $b$-$f $ mixture may become SUSY
\cite{yu}. We are interested in a special kind of molecule $b$, a
bound state of $f$ and another species of Fermi atom $F$ with
binding energy $E_{b}$, and restrict our analysis to the normal
phase of the $b$-$f$ mixture~\cite{note}. To probe the SUSY
behaviors, we load a free Fermi atom $F$ gas, which does not
interact both $b$ and $f$ directly. In a photoassociation(PA)
process~\cite{pa}, the transitions between two atoms and one
molecule, i.e., $Ff\leftrightarrow b$, are induced by two laser
beams with frequencies $\omega _{1}$ and $\omega _{2}$. For the
SUSY $b$-$f$ mixture, this resembles a high energy physics
process: a `quark' or a `lepton' ($f$) absorbs a `fermionic
gaugino' (`absorbs' a $F$ and emits a photon) and decays to a
`squark' or a `slepton' ($b$) or vice verse. (One can also
consider $f$ to be a Fermi molecule formed by the bound state of a
Bose atom $b$ and a Fermi atom $F$, i.e., processes
$Fb\leftrightarrow f$. We will study these processes separately.)

For a negative detuning $\delta _{0}=\omega _{2}-\omega
_{1}-E_{b}$, we show that the molecule dissociation process
$b\rightarrow Ff$ is forbidden. In the formation process
$Ff\rightarrow b$, two types of new fermionic excitations, an
individual (bosonic) particle-(fermionic) hole pair continuum and
a collective mode, emerge when the SUSY in the $b$-$f$ mixture is
slightly broken. For a SUSY $b$-$f$ mixture, the former is
completely suppressed while the latter in zero-momentum becomes an
exact eigenstate, the Goldstino mode \cite{yu}. In this sense, we
regard these excitations as {\it the SUSY responses}. The PA
spectrum is directly related to the the molecular formation rate
varying as the detuning and faithfully describes these two types
of excitations. The position of peak in the PA spectrum determines
the frequency of the collective zero-momentum mode. This molecular
formation rate is measured by the number variation of the $F$
atoms in time.  Experimentally, the number counting of
 atoms is much simpler than detecting the single atom
spectrum.

\vspace{2mm}

\noindent \textit{Model setup ---} The system illustrated in Fig.~\ref{fig1}%
(a) is described by a Hamiltonian $H=H_{0}+H_{ex}$, where $%
H_{0}=H_{bf}+H_{F} $ with $H_{bf}=H_{b}+H_{f}+V$. By means of the Feshbach
resonance \cite{Fech}, the scattering lengths between $F$ and the $b$-$f$
mixture can be adjusted to negligibly small. In the tight-binding
approximation, one has%
\begin{eqnarray}
H_{\alpha } &=&-\sum_{\left\langle ij\right\rangle }t_{\alpha }a_{i}^{\alpha
\dagger }a_{j}^{\alpha }-\mu _{\alpha }\sum_{i}a_{i}^{\alpha \dagger
}a_{i}^{\alpha }, \\
V &=&\frac{U_{bb}}{2}\sum_{i}n_{i}^{b}(n_{i}^{b}-1)+U_{bf}%
\sum_{i}n_{i}^{b}n_{i}^{f},  \notag
\end{eqnarray}%
where $a_{i}^{\alpha }=b_{i}$, $f_{i}$, and $F_{i}$ ($\alpha =b$, $f$, and $%
F $) are the annihilation operators of $b$, $f$ and $F$ at site $i$; $\mu
_{\alpha }$ and $n_{i}^{\alpha }=a_{i}^{\alpha \dag }a_{i}^{\alpha }$ are
chemical potentials and the number operators at site $i$. The definitions of
the hopping amplitudes $t_{\alpha }$ and the interaction strengths $%
U_{\alpha \beta }$ by the Wannier function $w_{\alpha
}(\mathbf{r})$ can be found in the literature \cite{bfm}. The
spatial inhomogeneity of optical lattices trapping the atoms and
molecules has been omitted. For the
subsystem $b$-$f$ mixture, the Hamiltonian $H_{bf}$ is SUSY invariant if $%
t_{b}=t_{f}$, $U_{bb}=U_{bf}$ and $\mu _{b}=\mu _{f}$ \cite{th2,yu}. In
order to prevent the phase separation, the parameters obey, e.g., $4\pi
t_{f}\rho _{f}U_{bb}>U_{bf}^{2}$ in 2-dimensions, where $\rho _{f}$ is the
density of $f$ atoms \cite{ps}. We choose the parameters of the system
obeying this condition.

The PA processes are realized by simultaneously shining two laser
beams with frequencies $\omega _{1}$ and $\omega _{2}$ into the
lattice (shown in Fig.~\ref{fig1}(a)). The $\omega_1$-beam may
turn two free atoms $f$ and $F$ into a higher energy bound state
$|1\rangle$ which then may transit to the molecule $b$ by emitting
a photon with frequency $\omega_2$. Meanwhile, the molecule $b$
may also be excited to $|1\rangle$ by the $\omega_2$-beam and
then is unbound with some probability by emitting a photon with frequency $%
\omega_1$. For large detuning $\Delta _{0}$, the state $\left\vert
1\right\rangle $ can be eliminated adiabatically, so that the PA
is modeled by the tight-binding Hamiltonian
\begin{equation}
H_{ex}=\sum_{i}(g_{i}b_{i}^{\dagger }f_{i}F_{i}e^{i\delta _{0}t}+\mathrm{H.c.%
}),
\end{equation}%
where the detuning $\delta _{0}=\omega _{c}-E_{b}$ with the effective driven
frequency $\omega _{c}=\omega _{2}-\omega _{1}$, and $g_{j}=g_{0}\exp (-i%
\mathbf{k}_{0}\cdot \mathbf{r}_{j})$ with $g_{0}\varpropto \int d^{d}\mathbf{%
r}\exp (-i\mathbf{k}_{0}\cdot \mathbf{r})w_{b}^{\ast }(\mathbf{r})w_{f}(%
\mathbf{r})w_{F}(\mathbf{r})$ being the coupling intensity independent of
the site.

In the $\mathbf{k}$-space, the Hamiltonian $H_{ex}$ is rewritten as%
\begin{equation}
H_{ex}=g_{0}\sqrt{\rho }(\sum_{\mathbf{k}}Q_{\mathbf{k-k}_{0}}^{\dagger }F_{%
\mathbf{k}}e^{i\delta _{0}t}+\mathrm{H.c.}),
\end{equation}%
where $Q_{\mathbf{k}}^{\dagger }=\sum_{\mathbf{p}}b_{\mathbf{p+k}}^{\dagger
}f_{\mathbf{p}}/\sqrt{N}$ and $a_{\mathbf{k}}^{\alpha }= \sum_{j}a_{j}^{\alpha }\exp (-i\mathbf{k}%
\cdot {\bf r}_j)/\sqrt{\mathcal{V}}$ (where $\mathcal{V}$ stands
for the volume and
$a_{\mathbf{k}}^\alpha $ stand for $b_{\mathbf{k}}$, $f_{%
\mathbf{k}}$, and $F_{\mathbf{k}})$. $\rho =N/%
\mathcal{V}$ is the total density of $b$ and $f$ with the particle number $%
N=\sum_{i}(n_{i}^{b}+n_{i}^{f})$.

\vspace{2mm}

\noindent \textit{Molecular formation rate ---} The formation rate of the
molecules $b$ can be counted by the PA variation of $F$-fermion number $%
R=\partial _{t}\left\langle \psi (t)\right\vert N_{F}\left\vert \psi
(t)\right\rangle $ for $\left\vert \psi (t)\right\rangle $ being the time
evolution from the ground state $\left\vert G\right\rangle =\left\vert
g\right\rangle \left\vert F\right\rangle $ of $H_{0}$. It follows from the
linear response theory \ that
\begin{equation}
R=2g_{0}^{2}\rho \sum_{\mathbf{k}}\text{Im}D_{R}(\mathbf{k},-\delta _{0}),
\label{R}
\end{equation}%
where the retarded Green function is given by%
\begin{equation}
D_{R}(\mathbf{k},\omega )=\int_{-\infty }^{\infty }dxA(\mathbf{k-k}_{0},x)%
\frac{n_{f}(x)-n_{f}(\varepsilon _{\mathbf{k}}^{F})}{x-\varepsilon _{\mathbf{%
k}}^{F}-\omega -i0^{+}},  \label{DT}
\end{equation}%
 in terms of one loop calculations (Fig.~\ref{fig1}(b)). The single particle
dispersions are $\varepsilon _{\mathbf{k}}^{\alpha }=-2t_{\alpha
}\sum_{s=1}^{d}\cos k_{s}-\mu _{\alpha }$ where the lattice
spacings are set to be the unit. \ $n_{f}(x)$ is the Fermi
distribution at temperature $T$ and the spectral function $A(\mathbf{k}%
,\omega )=-\mathrm{Im}\Pi _{R}(\mathbf{k},\omega )/\pi $ is defined by the
retarded Green function%
$
\Pi _{R}(\mathbf{k},\omega )=-i\int_{0}^{\infty }dt\left\langle g\right\vert
\{Q_{\mathbf{k}}(t),Q_{\mathbf{k}}^{\dagger }(0)\}\left\vert g\right\rangle
e^{i\omega t}.
$

At sufficiently low temperature, the pole and branch cut in
Eq.~(\ref{DT}) are not qualitatively affected by $T$ and nor is
the molecular formation rate. For simplicity, we take a zero
temperature approximation in our calculation. It follows from
Eq.~(\ref{DT}) that the rate $R=R_{b\rightarrow
Ff}-R_{Ff\rightarrow b}$ contains two parts :
\begin{eqnarray}
R_{b\rightarrow Ff} &=&\sum_{\mathbf{k}}2\pi g_{0}^{2}\rho A(\mathbf{k-k}%
_{0},\varepsilon _{\mathbf{k}}^{F}-\delta _{0})\theta (\delta
_{0}-\varepsilon _{\mathbf{k}}^{F}),  \notag \\
R_{Ff\rightarrow b} &=&\sum_{\mathbf{k}}2\pi g_{0}^{2}\rho A(\mathbf{k-k}%
_{0},\varepsilon _{\mathbf{k}}^{F}-\delta _{0})\theta (-\varepsilon _{%
\mathbf{k}}^{F}).  \label{8}
\end{eqnarray}%
which respectively are the dissociation rate for $b\to Ff$ and the formation
rate for $Ff\to b$.

\vspace{2mm}

\noindent \textit{Collective and individual fermionic modes }--- In order to
obtain $R $ for the weak interactions, we perturbatively calculate $\Pi _{R}(%
\mathbf{k},\omega )=\rho ^{-1}[\Pi _{0}^{-1}(\mathbf{k},\omega
)+U_{bf}]^{-1} $, which formally results from the equation of motion of $Q_{%
\mathbf{k}}.$ It then follows from the random phase approximation (RPA) \
illustrated by the \textquotedblleft bubble\textquotedblright in (Fig. \ref%
{fig1}(b)) that
\begin{equation}
\Pi _{0}(\mathbf{k},\omega )=\int \frac{d^{d}\mathbf{p}}{(2\pi )^{d}}\frac{%
n_{f}(\varepsilon _{\mathbf{p}}^{f})+n_{b}(\varepsilon _{\mathbf{k+p}}^{b})}{%
\omega -E_{\mathbf{kp}}+i0^{+}},
\end{equation}%
where $
E_{\mathbf{kp}}=\varepsilon _{\mathbf{k+p}}^{b}-\varepsilon _{\mathbf{p}%
}^{f}+2\rho _{b}\delta U+U_{bf}\rho$ with $\delta U=U_{bb}-U_{bf}$
and $\rho _{b}=N_{b}/\mathcal{V}$; $n_{b}(x)$
is the Bose distribution. The isolated pole and branch cut of $\Pi _{R}(%
\mathbf{k},\omega )$ describe the collective and individual SUSY excitations
of $Q_{\mathbf{k}}^{\dagger }\left\vert g\right\rangle $.

\begin{figure}[tbp]
\includegraphics[bb=25 194 581 657, width=8.5 cm, clip]{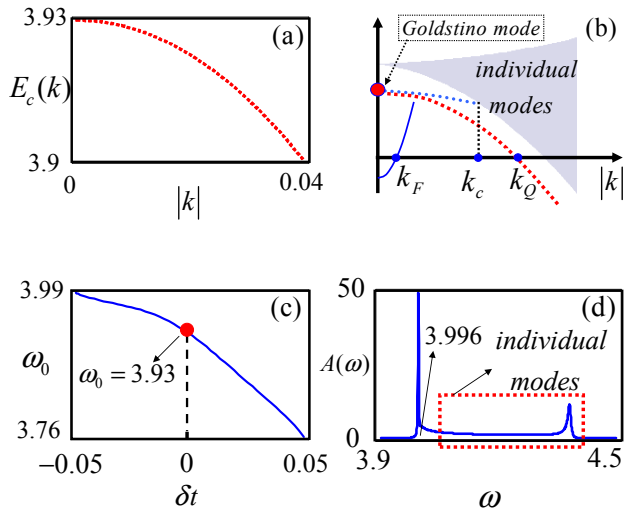}
\caption{(Color online) The dispersions, spectrum and spectral function of
the excitations for $\Delta \protect\mu=3.93$ and $U_{bb}=U_{bf}=0.1$; $%
t_{b} $ is taken as the unit. (a) The dispersion of collective mode for the
SUSY mixture. (b) The spectrum for the SUSY system: the dashed (red) and
dotted (blue) curves denote the dispersions of collective modes for $\protect%
\theta=0$ and $\protect\pi/5$, respectively. The solid (blue)
curve denotes the dispersion of atoms $F$ for a small Fermi
momentum $k_{F}$. (c) The
frequencies of $Q_{0}^{\dagger }\left\vert g\right\rangle $ for different $%
\protect\delta t$. (d) The spectral function of the zero-momentum for $%
\protect\delta t=-0.1$: The sharp peak is the shifted
goldstino-like mode.}
 \label{fig2}
\end{figure}

Next we consider the elementary excitations in the two-dimensional
lattice with $f$ atoms at half filling, i.e., $\rho _{f}=0.5$. For
the SUSY $b$-$f$ mixture, i.e., $\delta U=0$ and $\delta
t=t_{b}-t_{f}=0$, the dispersion of the collective modes,
$E_{c}(\mathbf{k})\simeq \Delta \mu -\alpha \left\vert
\mathbf{k}\right\vert ^{2}$ for the small $\left\vert \mathbf{k}\right\vert $%
~\cite{yu}, is read out from the poles of the retarded Green function $\Pi
_{R}(\mathbf{k},\omega )$ (see Fig.~\ref{fig2}(a)), where $\Delta \mu =\mu
_{f}-\mu _{b}$. For large $\left\vert \mathbf{k}\right\vert $, the energy $%
E_{c}(\mathbf{k})=E_{c}(\left\vert \mathbf{k}\right\vert ,\theta )$ depends
not only on $\left\vert \mathbf{k}\right\vert $ but also on the angle $%
\theta =\arctan (k_{y}/k_{x})$. The energy $E_{c}(\left\vert \mathbf{k}%
\right\vert ,\theta )$ of $Q_{\mathbf{k}}^{\dagger }\left\vert
g\right\rangle $ decreases as $\left\vert \mathbf{k}\right\vert $
increases for a fixed $\theta $. The retarded Green's function
$\Pi _{R}(0,\omega )=(\omega -\Delta \mu +i0^{+})^{-1}$ possesses
a pole $\omega =\Delta \mu $, which corresponds to the
goldstino-like excitation $Q_{0}^{\dagger }\left\vert
g\right\rangle $. This recovers the result in Ref.~\cite{yu}. The
excitation spectrum is schematically shown in Fig.~\ref{fig2}(b).

For the $b$-$f$ mixture deviating slightly from SUSY, the retarded Green
function $\Pi _{R}(0,\omega )$ has an isolated pole and a branch cut, which
correspond to \textit{a collective fermionic mode and individual (bosonic)
particle-(fermionic) hole pair continuum modes}, respectively. The pole in $%
\omega _{0}<E_{0}-4\delta t$ for $\delta t>0$ (or $\omega _{0}<E_{0}$ for $%
\delta t<0$) describes the shifted goldstino-like mode. The frequencies $%
\omega _{0}$ of collective zero-momentum mode for different $\delta t$ are
shown in Fig.~\ref{fig2}(c). For $\mathbf{k}\neq 0$, the pole of $\Pi _{R}(%
\mathbf{k},\omega )$ has the form $E_{c}^{\prime }(\mathbf{k})\simeq \omega
_{0}-\alpha ^{\prime }\left\vert \mathbf{k}\right\vert ^{2}$ for small $%
\left\vert \mathbf{k}\right\vert $. Remarkably, the branch cut $l_{0}$\ of $%
\Pi _{R}(0,\omega )$ emerges, which describes individual zero-momentum
modes. Here, $l_{0}=[E_{0}-4\delta t,E_{0}]$ for $\delta t>0$ (or $%
[E_{0},E_{0}-4\delta t]$ for $\delta t<0$), and $E_{0}=\Delta \mu
+2\rho _{b}\delta U+U_{bf}\rho $. Notice that for the weak
interactions $U_{bb}$ and $U_{bf}$ the SUSY breaking from $\delta
U$ does not develop a branch cut but only shifts the positions of
the pole and branch cut. The pole and the branch cut can be seen
in the spectral function $A(\mathbf{k},\omega )$, i.e., the peak
and the hump in Fig.~\ref{fig2}(d) for $\mathbf{k=0}$.  Note that
for the SUSY $b$-$f$, the branch cut length $l_0$ of $\Pi
_{R}(0,\omega )$ shrinks to zero so that the individual continuum
modes of zero momentum are completely suppressed. On the other
hand, as the $b$-$f$ mixture deviates from the SUSY, the
goldstino-like mode is gradually suppressed. We examine the
dependence of the spectral function on the interacting strength
and find that the hump height may be depressed as the interaction
becomes stronger, e.g., the height is lower than 0.5 for
$U_{bf}=U_{bb}=0.5$ comparing with $\sim 10$ in Fig.~\ref{fig2}(d)
for $U_{bf}=U_{bb}=0.1$.

In order to study the PA spectrum of the molecular formation rate, we
discuss the excitation spectrum shown in Fig.~\ref{fig2}(b). For some
momenta $\mathbf{k}$, there is a collective mode (dashed red curve) below
the individual continuum. For other momenta $\mathbf{k}_{c}$, the dispersion
of the collective mode merges into the continuum. However, for small
momentum $\mathbf{k}$, there always exists a collective mode below the
individual continuum. For convenience, we define a critical momentum $%
k_{Q}(\theta )$, so that for a fixed $\theta $, when $\left\vert \mathbf{k}%
\right\vert >k_{Q}(\theta )$ the negative frequencies of the mode $Q_{%
\mathbf{k}}$ emerge, i.e., $A(\mathbf{k},\omega <0)\neq 0$ when $\left\vert
\mathbf{k}\right\vert >k_{Q}(\theta )$, and $A(\mathbf{k},\omega <0)=0$ when
$\left\vert \mathbf{k}\right\vert <k_{Q}(\theta )$.

\vspace{2mm}

\noindent \textit{PA spectrum ---} The rate $R$ varies as detuning
$\delta _{0}$ or the light frequency $\omega _{c}$. Measurement of
the $b$ boson formation rate varying as $\delta _{0}$ is called
the PA spectrum $S(\delta _{0})$. For a long wave photon, the
coupling $g_{j}$ varies slowly  in space and the rates in
Eq.~(\ref{8}) are approximately independent of $\mathbf{k}_{0}$.

According to Eq.~(\ref{8}), the dissociation rate $R_{b\rightarrow Ff}$ does
not vanish only if $\delta _{0}>\varepsilon _{\mathbf{k}}^{F}$ and $A(%
\mathbf{k},\varepsilon _{\mathbf{k}}^{F}-\delta _{0})\neq 0$.
Because the spectral function $A(\mathbf{k},x<0)\neq 0$ is defined
by the retarded Green function, it does not vanish only when the
energies for collective modes
or individual modes of $Q_{\mathbf{k}}$ are negative for the large $|\mathbf{%
k}|>k_{Q}(\theta )$. We consider a dilute Fermi gas $F$ with the chemical
potential $\mu _{F}\sim -4t_{F}$, the dispersion relation turns out $%
\varepsilon _{\mathbf{k}}^{F}=t_{F}\left\vert \mathbf{k}\right\vert ^{2}-\mu
_{\mathrm{eff}}$, where $\mu _{\mathrm{eff}}=\mu _{F}+4t_{F}$. In this case,
the fermion $F$ possesses a small Fermi momentum $k_{F}=\sqrt{\mu _{\mathrm{%
eff}}/t_{F}}$ which is much smaller than $k_{Q}(\theta )$ for small
deviating $\delta t$. Therefore, $\varepsilon _{k}^{F}$ is\ always positive
when $\left\vert \mathbf{k}\right\vert >k_{Q}$ (see Fig.~\ref{fig2}(b)). For
the negative detuning $\delta _{0}$, the condition $\delta _{0}>\varepsilon
_{k}^{F}$ is not satisfied in the regime $\left\vert \mathbf{k}\right\vert
>k_{Q}(\theta )$. That is, $A(\mathbf{k},\varepsilon _{\mathbf{k}%
}^{F}-\delta _{0})$ and $\theta (\delta _{0}-\varepsilon _{k}^{F})$ can not
be non-zero simultaneously for the negative detuning and small Fermi
momentum $k_{F}$. This finishes our proof of $R_{b\rightarrow Ff}=0$.

The vanishing of $R_{b\rightarrow Ff}$ for the negative $\delta
_{0}$ and small Fermi momentum $k_{F}$ can be understood in a more
straightforward way. The
transition $b\rightarrow Ff$ is described by the Hermite conjugate term ($%
\mathrm{H.c.}$) in the Hamiltonian $H_{ex}$, which is a
high-frequency oscillation term when $\delta _{0}<0$. Hence, the
Fermi golden rule results in that $R_{b\rightarrow Ff}$ vanishing
under the first-order perturbation (linear response).

\begin{figure}[tbp]
\includegraphics[bb=33 424 572 631, width=8.5 cm, clip]{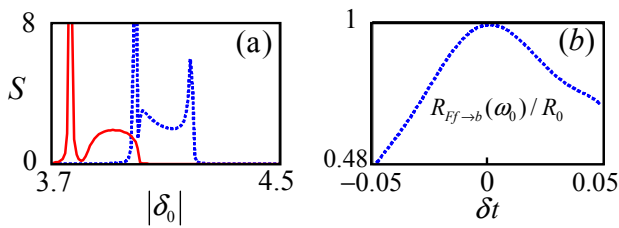}
\caption{(Color online) The PA spectra with the same
parameters as those in Fig. 2. (a) The detuning dependence for $\protect%
\delta t=0.05$ and $-0.05$, the solid (red) and dashed (blue) lines,
respectively. The unit of $S$ is $2\protect\pi g_{0}^{2}\protect\rho N_{F}$
and $N_{F}/N$ is taken to be 0.01. (b) The major peak values for different $%
\protect\delta t$. This shows that the peak value is suppressed when the
system deviates from SUSY.}
\label{fig3}
\end{figure}

For the negative $\delta _{0}$ and small Fermi momentum $%
k_{F}$ ($\mu _{\mathrm{eff}}\ll t_{F}$), the molecule formation rate now is
reduced to $R=-R_{Ff\rightarrow b}$ and%
\begin{equation*}
R_{Ff\rightarrow b}\simeq \Biggl\{%
\begin{array}{l}
Z_{0}g_{0}^{2}N/[2(t_{F}+\alpha ^{\prime })],~~~\mathrm{for}~\delta
_{0}=-\omega _{0}, \\
2\pi g_{0}^{2}\rho N_{F}A(0,-\delta _{0}),~~\mathrm{for}~\left\vert \delta
_{0}\right\vert \in l_{0},\label{12} \\
0,~~~~~~~~~~~~~~~~~~~~~~~~~~\mathrm{otherwise}%
\end{array}%
\end{equation*}%
which leads to our main result: \textit{The PA spectrum $\
S(\delta _{0})=-R_{Ff\rightarrow b}$ (see Fig.~\ref{fig3}(a))
displays the spectral function of excitations} $Q_{0}^{\dagger
}\left\vert g\right\rangle $.

For the SUSY $b$-$f$ mixture, the length of branch cut
$l_{\mathbf{0}}$ tends to zero and the individual modes are
suppressed. Meanwhile, the residue $Z_{0}=1$ and the formation
rate $R_{Ff\rightarrow b}=g_{0}^{2}N/[2(t_{F}+\alpha )]\equiv
R_{0}\varpropto N$ at $\delta _{0}=-\Delta \mu $, and vanishes for
the other detunings. There is a sharp peak at $\delta _{0}=-\Delta
\mu $ in the PA spectrum. For a generic $b$-$f$ mixture deviating
from SUSY, the residue $Z_{0}<1$ decreases as $|\delta t|$
increases. As a result, the peak height is lowered while its
position is shifted to $\delta _{0}=-\omega _{0}$. The ratio
$R_{Ff\rightarrow b}(\omega _{0})/R_{0}$ is shown in
Fig.~\ref{fig3}(b) for different $\delta t$ and a small $\mu
_{\mathrm{eff}}$, where $R_{Ff\rightarrow b}(\omega _{0})$ is the
value of $R_{Ff\rightarrow b}$ at $\delta _{0}=-\omega _{0}$.
Remarkably, a minor hump develops in the region $\delta _{0}\in
l_{0}$ due to the emergence of individual modes (See
Fig.~\ref{fig3}(a)). As the system deviates further from SUSY,
 the individual modes are enhanced due to the sum rules
 $\int d\omega A(\mathbf{0},\omega )=1$. The temperature may suppress and broaden
the peak and hump. These characters of the PA spectrum in the
$b$-$f $ mixture are experimentally measurable SUSY responses to
the light field.

\noindent \textit{Conclusions }--- We studied how to observe the
SUSY nature of the $b$-$f$ mixture in optical lattices through PA
spectra. For the Bose molecules formed with two species of Fermi
atoms, we showed that the photon induced atom-molecule transition
displays the signal of SUSY. As the response to the PA processes,
a fermionic individual continuum and the goldstino-like mode were
found. The PA spectrum can explicitly witness the molecular
formation rate of $Ff\to b$. Because the goldstino-like mode in
zero momentum turns to be the exact eigen state for the SUSY
mixture, the major peak in the PA spectrum reflects the SUSY
response to the light field, even if the mixture is not fine-tuned
to a SUSY one.

The authors thank Jinbin Li and Peng Zhang for the useful
discussions. YY is grateful to Kun Yang for sharing his idea in
the earlier stage of this work. This work is supported in part by
National Natural Science Foundation of China, the national program
for basic research of MOST of China and a fund from CAS.



\begin{thebibliography}{99}
\bibitem{note} If the bosons $b$ are condensed, the elementary excitations
are more fruitful.  We will leave them in further studies.
\end{thebibliography}
\end{document}